\documentstyle[aps,prl]{revtex}
\begin{document}
\draft 

\title{Quasiparticle transport in the vortex state of d-wave superconductors.}
\author{I. Vekhter$^1$, J. P. Carbotte$^2$, E. J. Nicol$^1$}
\address{${}^1$Department of Physics, University of Guelph, Guelph, 
	Ontario N1G 2W1, Canada\\
	${}^2$Department of Physics and Astronomy, McMaster University,
	Hamilton, Ontario  L8S 4M1,  Canada}
\date{\today}
\maketitle

\begin{abstract}
We calculate the magnetic field and temperature dependence of the 
Raman response, superfluid density and the NMR relaxation rate 
in the vortex state of a d-wave superconductor arising from the 
Doppler energy shift of extended quasiparticle states. 
Our results are valid both at low temperatures, where
 we observe scaling with variable $TH^{-1/2}$ and obtain explicit 
form of the scaling functions,  and beyond this region. 
We derive a universal frequency dependent scaling relation for the 
Raman response, and discuss the breakdown of the single relaxation 
rate approach to NMR response.
\end{abstract}


In the last few years there has emerged a consensus regarding the d-wave
symmetry of the 
order parameter in the hole doped high-$T_c$ cuprates.
While in conventional, s-wave, superconductors a finite energy gap exists 
everywhere on the Fermi surface, the d-wave order parameter  has lines of 
nodes, 
which leads to a gapless excitation spectrum along certain directions in 
momentum space. 
Consequently the low
temperature behavior 
of the thermodynamic and transport coefficients 
in the high-$T_c$ materials are qualitatively different from those of the 
s-wave compounds.
Properties of
 the vortex state in ``unconventional'' superconductors also
differ significantly 
from those obtained within the framework of s-wave superconductivity,
and a
clear understanding of these properties is essential for  
interpretation of the experimental results and a
better grasp of the novel physics associated with the superconducting state of
 the high-$T_c$ materials.

Volovik \cite{Vol2,Kop}
pointed out that while in an s-wave superconductor the density of states 
is determined at 
low fields  by the quasiparticles in the vortex core, both the 
density of states and the entropy in 
d-wave superconductors are dominated by the extended quasiparticle states
which exist even at zero temperature in the nodal directions of the 
order parameter. A remarkable consequence of this behavior is
that the specific heat of such a superconductor 
varies as $\sqrt H$ rather than 
linearly in the applied field. Other authors \cite{Simon}
used the Dirac form of the low-energy 
excitation spectrum of nodal quasiparticles
to demonstrate 
that thermal and transport coefficients exhibit scaling with $TH^{-1/2}$.
However, strictly speaking, the analysis applies 
only to clean superconductors, and the energy spectrum is only 
Dirac-like at energies small compared to 
the 
gap amplitude $\Delta_0$; a crossover to a different scaling
regime followed by the breakdown of scaling
have been predicted 
at $(T/T_c)(H_{c2}/H)^{1/2}\sim 1$\cite{Kop,Simon}.

Very recently K\"ubert and Hirschfeld\cite{Hirsch1,Hirsch2}
placed these scaling arguments in the 
framework of the Green's function formalism capable of treating both 
the energies of order of the gap and the effects of disorder. 
These authors 
argued that for a short coherence length superconductor the typical  spacing 
of the energy levels in the core, $\Delta_0^2/E_f$, where $E_f$ is the Fermi 
energy, is large so that only one or a few states (if any) exist there, 
and suggested that the contribution of the vortex cores 
to the transport coefficients is negligible over a 
wide region of
$H$ and $T$. 
They proposed to  
 account for the effect of the magnetic field on the extended states 
semiclassically by 
introducing a Doppler shift due to  
circulating supercurrents, which
for $H\ll H_{c2}$ are approximated by  the superfluid velocity 
field
around a single vortex ${\bf v}_s=\hbar\hat\theta/2mr$, 
where $r$ is the distance from the center of the vortex and 
$\theta$ is the azimuthal angle in real space.
The authors of Ref.\cite{Hirsch2} 
investigated in detail the breakdown of scaling 
of both specific heat and the thermal conductivity 
with increased impurity scattering, and the results agreed 
remarkably well with recent experiments \cite{Tail}.

In this work we use the same approach 
to examine the effect of magnetic fields 
$H_{c1}\leq H \ll H_{c2}$
on 
the Raman response, superfluid density, Knight shift, and NMR relaxation rate 
in a d-wave superconductor. These are issues of considerable experimental 
interest:
recently Blumberg {\it et al.} \cite{blum1} 
analyzed 
the changes induced by a magnetic field  in the electronic part of 
the Raman response; 
NMR relaxation rates and the Knight shift are measured in fields of up to 10T
\cite{Take}, in the vortex state, while 
muon spin rotation ($\mu$SR) 
is used to determine the low temperature penetration depth in a magnetic 
field \cite{musr}, which is related to the superfluid density. 
In our analysis we obtain 
scaling similar to that suggested
in Ref. \cite{Simon,Kop} and give the explicit form of the scaling functions
at low temperatures, 
however our  results remain valid in a wider parameter range.
We do not include impurity scattering in the present calculations, 
if included, it affects only  the extreme low frequency 
(low temperature) part of the Raman (NMR) response, 
and the results presented 
here remain valid beyond this narrow region.

We employ the single particle Green's function 
which is
obtained by introducing the Doppler shift
into the BCS function \cite{Maki,Hirsch1}
\begin{equation}
\label{Green}
G({\bf k},\omega_n; {\bf r})=-
{(i\omega_n-{\bf v}_s{\bf k})\tau_0+\Delta_{\bf k}\tau_1+\zeta_{\bf k}\tau_3
\over
\omega_n^2+\zeta_{\bf k}^2+\Delta_{\bf k}^2},
\end{equation}
where $\omega_n$ is the fermionic Matsubara frequency, 
$\zeta_{\bf k}$ is the energy of a quasiparticle with momentum {\bf k},
measured 
with respect to the Fermi level, 
and $\tau_i$ are Pauli matrices. The Green's function depends on the coordinate
{\bf r} in real space via the superfluid velocity ${\bf v}_s$.
Now thermal and transport coefficients can be calculated 
using the 
standard approach \cite{AGD}, however, 
they become local quantities which have to be 
averaged over a unit cell of the vortex lattice. Following Ref.\cite{Hirsch1}
we approximate 
this unit cell by a circle of radius $R$, 
where $2R=\xi_0\sqrt{2\pi}a^{-1}(H_{c2}/H)^{1/2}$ 
is the intervortex spacing, and $a$ is a geometric constant of order unity, 
so that the average of a quantity $f({\bf r})$ is given by
\begin{equation}
\label{average}
f(H)={1\over \pi R^2} \int d^2r f(r,\theta).
\end{equation}
Here we consider a model cylindrical Fermi surface 
and an experimental arrangement with the magnetic 
field parallel to the axis of the cylinder, ${\bf H}\| c$.

\paragraph*{Raman response.}
The Raman intensity is proportional to the imaginary part of the 
zero momentum density-density correlation function
\begin{equation}
\label{Raman}
\chi(i\Omega_n)=
-\gamma_R^2T
\sum_{{\bf k},\omega_m}f_s^2({\widehat{\bf k}})
Tr\bigl[\tau_3 G({\bf k}, i\omega_m)\tau_3 
	G({\bf k}, i\omega_m-i\Omega_n)\bigr],
\end{equation}
where $\gamma_R f_s({\widehat{\bf k}})$ is the Raman vertex. 
In Eq.(\ref{Raman}) vertex corrections due to 
Coulomb screening have been ignored, 
they appear only in the fully symmetric channel ($A_{1g}$), 
while we are interested primarily in the $B_{1g}$ and 
$B_{2g}$ scattering geometries.
After doing the Matsubara sum and analytically 
continuing the response function to real frequencies, 
we obtain the local Raman response
\begin{equation}
\chi''(\Omega;{\bf r})={1\over 2}
\gamma_R^2 N(0)\int_{FS}d\widehat{\bf k} f_s^2(\widehat{\bf k}_f)
{1\over \Omega\sqrt{(\Omega/2)^2-\Delta_{\bf k}^2}}
\Bigl[ \tanh{\Omega-2{\bf v}_s\widehat{\bf k}_f\over 4T}
	+ \tanh{\Omega+2{\bf v}_s\widehat{\bf k}_f\over 4T}\Bigr],
\label{locRam}
\end{equation}
where the integration  is over the Fermi surface, 
and $k_f$ is the Fermi momentum.
Note that the kernel of the integral is identical to that in zero field, 
and that the field dependence is only in  the thermal factors. 
Indeed, as
the Doppler shift is the same 
for the quasiparticles absorbing and emitting photons
the difference in energy between the initial and the final states 
remains unchanged, whereas the thermal factors, which
depend on the local value of the chemical potential, 
are strongly affected by the magnetic field. 

To obtain the field dependent Raman intensity 
 we spatially average the local response given in 
Eq.(\ref{locRam}) according to Eq.(\ref{average}). 
A crucial observation is that since the 
spatial average is performed over all directions of the superfluid velocity, 
the result does not depend on specific position 
$\widehat{\bf k}_f$ at the Fermi surface. Consequently, spatial
averaging  decouples from the integration over the Fermi surface, 
and we arrive at the following universal 
scaling relation for the Raman intensity
\begin{equation}
\label{scaling}
\chi''(\Omega;H)=\chi''(\Omega;0)F({\Omega/2T}, {E_H/T})
/F({\Omega/2T}, 0),
\end{equation}
where $E_H=a\Delta_0\sqrt{H/H_{c2}}$ 
is the typical quasiparticle Doppler shift, which is the energy scale
introduced by the magnetic field, and 
\begin{equation}
\label{strictF}
F(x,y)= {1\over \pi}\int_0^1 zdz\int_0^{2\pi}d\theta
\tanh\bigl({x\over 2}- \sqrt{\pi\over 8}{y\over z}\cos\theta\bigr)
\end{equation}
is a generalization of the thermal function 
$F(\epsilon/T,0)=\tanh(\epsilon/2T)$.
In general the function $F$ has to be evaluated numerically, 
however it can be obtained analytically 
in the important
limit $T=0$ when it depends only on the ratio
$x/y$:
\begin{equation}
\label{scaleT0}
$$F_0(w)= \cases{1-1/(2w^2) & $w\ge1$;\cr
	\pi^{-1}\bigl[(2-w^{-2})\arcsin w +\sqrt{w^{-2}-1}\bigr], 
			& $w\le 1$, \cr}$$
\end{equation}
here $w=(2/\pi)^{1/2}(x/y)$. 
Then the  ratio of Raman intensities in 
{\it any} channel at low temperatures is given by
\begin{equation}
\label{ramant0}
{{\chi''}_{T=0}(\Omega;H)\over{\chi''}_{T=0}(\Omega;0)}=
F_0\bigl({\Omega\over \sqrt{2\pi}E_H}\bigr).
\end{equation}
The intensity is most strongly affected for the Raman shifts 
below the average Doppler shift $E_H$, where the  the scaling function $F_0$
is 
almost linear in $\Omega H^{-1/2}$, while for large frequencies
the field dependent correction is small and linear in the applied field.
In Fig.\ref{fig:f0} we also show that 
the main features of the scaling function remain robust at low temperatures.

So far we have made no assumptions regarding the 
specific symmetry of the order parameter or the particular Raman geometry. 
Assuming a $d_{x^2-y^2}$-wave symmetry, we obtain that 
the intensity at small frequencies $\Omega\ll E_H$ in the 
$B_{1g}$ and $B_{2g}$ channels become quartic and quadratic, respectively, 
compared to cubic and linear dependence in the absence of field. 
It also follows that the height of the $2\Delta_0$ peak in the 
$B_{2g}$ channel decreases linearly with the magnetic field. 
Finally the integrated normalized
Raman intensity 
$\int_0^\infty d\Omega [{\Delta\chi''(\Omega;H)/\chi''(\Omega;0)}]$
scales with $\sqrt H$,
while the integral of the change in the signal 
itself depends on the particular Raman geometry \cite{blum2}.

\paragraph*{Superfluid density.}

Using the Green's function given in Eq.(\ref{Green})
we obtain that 
the relative change in superfluid density 
is given by
the generalized Yosida function\cite{Yosida}
\begin{equation}
\label{Knight}
{\delta n_s(T,H)\over n_s(0,0)}=
{1\over 2 N(0)}\sum_{\bf k}{\partial\over\partial E_{\bf k}}
F \Bigl({E_{\bf k}\over T}, {E_H\over T}\Bigr),
\end{equation}
where $E_{\bf k}=\sqrt{\zeta_{\bf k}^2 +\Delta_{\bf k}^2}$, and the 
function F was defined in Eq.(\ref{strictF}).
It is clear from the zero-temperature limit $F_0$ of the thermal
function $F$, and from
Fig.\ref{fig:f0}, that at low 
temperatures $T\ll E_H$ the energy scale determining the range of 
$F(E_{\bf k}/T, E_H/T)$ as a function of $E_{\bf k}$ is the
magnetic energy $E_H$ rather  
than $T$, so that  the behavior of transport coefficients is 
drastically different from the $H=0$ case.
As we show in  Fig.\ref{fig:nsup},
while in absence of magnetic field the superfluid density
$n_s$ decreases linearly with $T$ due to the 
linear low-energy density of states of a superconductor 
with lines of nodes of the gap, the low temperature behavior in the applied
field becomes
\begin{equation}
\label{lowTknight}
{\delta n_s(T,H)\over n_s(0,0)}
\approx \sqrt{8\over \pi}{E_H\over \Delta_0}+
{2\over 9}{\sqrt{2\pi}T^2\over E_H \Delta_0},
\end{equation}
and the temperature dependent term shows scaling with $T^2H^{-1/2}$.
In all the numerical work we have used the BCS value
$\Delta_0=2.14T_c$.
The superfluid density can be extracted from either optical conductivity
in the magnetic field, which has not been measured yet, or from the
 $\mu SR$
 experiments, which
determine the London penetration depth $\lambda(H)$ as $T\rightarrow 0$. 
The exact relationship between the penetration depth 
and the superfluid density in the mixed state is not yet clearly understood, 
since nonlinear\cite{Yip} and nonlocal \cite{Leggett} effects are believed 
to be important, however, 
the non-linear, in $H$, behaviour similar to that
given in Eq.(\ref{lowTknight})
has been obtained 
numerically when both 
of these effects are included \cite{Marcel}.
The inset of Fig.\ref{fig:nsup} shows the data of Ref.\cite{musr} 
for a three-crystal mosaic of $YBCO$ with our best linear, in $\sqrt H$,
fit, which gives $\lambda(0)=1142{\rm\AA}$ and $H_{c2}/a^2=88T$, 
compared to  $\lambda(0)=1155{\rm\AA}$ used in Ref.\cite{musr} 
for a linear  in $H$ fit.  Because of nonlinear effects the obtained
value for $H_{c2}$ is a low-end estimate
for the upper critical field.

The scaling behavior given by Eq.(\ref{lowTknight}),
is shown in Fig.\ref{fig:scaleKnight} where  
the numerical results follow the scaling curve up to 
$T/E_H\approx(T/\Delta_0)(H_{c2}/H)^{1/2}\sim 1$.
For larger $T/E_H$ results obtained for small $E_H$ and, consequently,
low temperatures still scale, while for larger fields the scaling is broken
since corresponding temperatures become high.

We note that in conventional linear response theory 
the relative change to the spin part of the Knight shift 
in fields $\mu_B H\ll T$ is 
also given by
Eq.(\ref{Knight}). However, in the present problem
there is an additional
magnetization due to the vortex lattice, which
has to be computed
from the grand potential\cite{won}. This contribution is small in
the temperature dominated regime
$T\gg E_H$, while at low temperatures
$M\approx (a^2\pi^2 T^2 \Delta_0 N(0))/(6 E_H H_{c2})$.

\paragraph*{Spin-lattice relaxation rate.}
Short-range antiferromagnetic correlations, which 
are believed to be important for the Cu
NMR relaxation rate in the cuprates,
cancel on the oxygen sites, 
resulting in a normal Korringa behavior seen in experiment\cite{Take,Hammel}. 
Then the relaxation rate $T_1^{-1}$ 
can be calculated using the low frequency 
limit of the uniform spin susceptibility, yielding 
\begin{equation}
\label{T1r}
{1\over T_1(r,\theta)T}={1\over 2T_1^{(c)}T_c}\int_{-\infty}^{+\infty}
N^2(E){\partial\over\partial E} \tanh{E-{\bf v}_s{\bf k}\over 2T},
\end{equation}
where  $T_1^{(c)}$ is the relaxation time at $T_c$, and
 $N(E)$ is the superconducting density of states.
It is important to note that
the NMR experiments measure the decay of magnetization
${\bf M}(t)\propto \exp(-t/T_1)$, rather than $T_1$ directly, 
so that it is  ${\bf M}(t)$ that has to be spatially averaged. 
For a distribution of $T_1({\bf r})$ the average magnetization 
cannot be described by a single relaxation rate, and, if such a fit is made,
the obtained value of $T_1$ is different depending on whether short or 
long time scale behavior is analyzed. At low temperatures $T\ll T_c$, 
if $T\geq E_H$, 
the times over which measurements are done, $t\sim T_1$,
correspond to  short time scale in the field-dependent term, and the
magnetization decay approximately
follows a single relaxation rate behavior with
\begin{equation}
\label{T1av}
{T_cT_1^{(c)}\over T_1T}=
{\pi^2\over 3}{T^2\over\Delta_0^2}
+{\pi\over 2}{E_H^2\over\Delta_0^2}\ln\Bigl[{T\over E_H}\Bigr].
\end{equation}
On the other hand, for $T_c\gg E_H\gg T$, 
the single relaxation time picture breaks down 
completely due to strong spatial variations of $T_1$. Even though
${\bf M}(t)$ is still described by the approximate relaxation rate
given by Eq.(\ref{T1av}) for $t\ll T_1$, in the experimentally relevant region
${\bf M}(t)\propto \exp(-t/T_1^{(H=0)})/t^{1/2}$.
We are aware that an independent analysis of NMR magnetization
data, including a wider range of temperatures and fields, is
being carried out, using an approach similar in spirit to this
one\cite{Rachel}. 

To conclude we have 
presented an approach to the calculation 
of thermal and transport properties of 
d-wave
 superconductors in the mixed state over 
a wide range of temperatures and fields, considering
the contribution of the extended quasiparticle states.
We obtained the explicit form of the scaling functions
in the low temperature regime $T\leq E_H$, 
and observed the breakdown of scaling at higher temperatures.
Our results agree qualitatively with the $\mu SR$ measurements
of the penetration
depth in the vortex state.

We are grateful to G. Blumberg, P.J. Hirschfeld and C. K\"ubert 
for important discussions, and we thank R. Wortis for discussing her
results before publication.
This research 
has been supported in part by NSERC of Canada (EJN and JPC) and CIAR (JPC).
EJN is a 
Cottrell Scholar of Research Corporation.

\begin{figure}
\caption{Zero temperature scaling function $F_0(w)$ (solid line). The result
of numerical evaluation of the function $F$, given in Eq.(\ref{strictF}), 
is plotted for $T=0.03\Delta_0$ for comparison (dashed line).}
\label{fig:f0}
\end{figure}

\begin{figure}
\caption{Superfluid density $n_s$ as a function of reduced 
temperature for different magnetic fields:
$E_H=0$ (solid line), $E_H=0.1T_c$ (dashed line), $0.2T_c$ (long-dashed line),
$0.3T_c$ (dot-dashed line). Inset: zero-temperature
penetration depth
data from $\mu$SR experiment plotted vs. $\sqrt H$.
Solid line: best linear fit. Slope corresponds to $H_{c2}/a^2=88T$.}
\label{fig:nsup}
\end{figure}

\begin{figure}
\caption{Full numerical evaluation of the superfluid density from 
Eq.(\ref{Knight}). Dashed line: low temperature result from 
Eq.(\ref{lowTknight}).}
\label{fig:scaleKnight}
\end{figure}

\stop
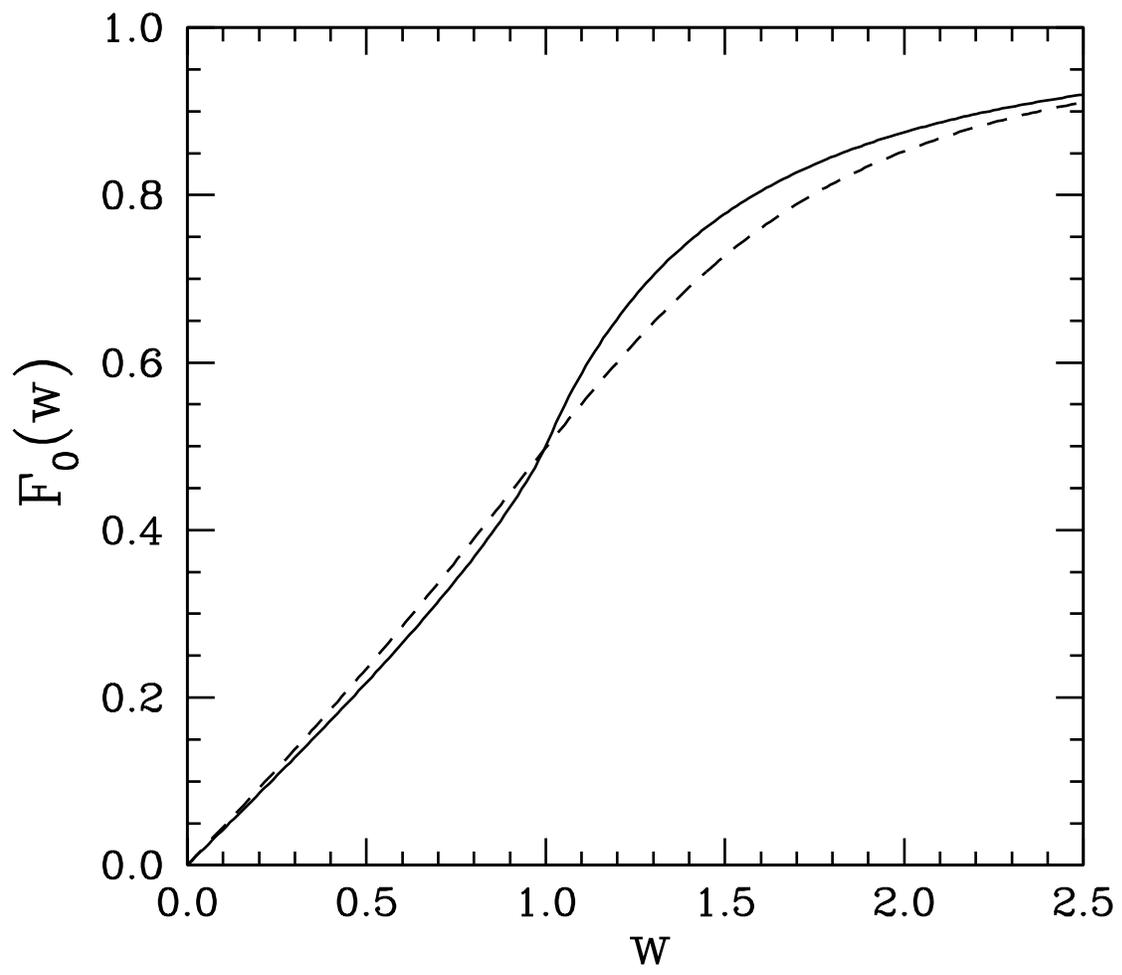
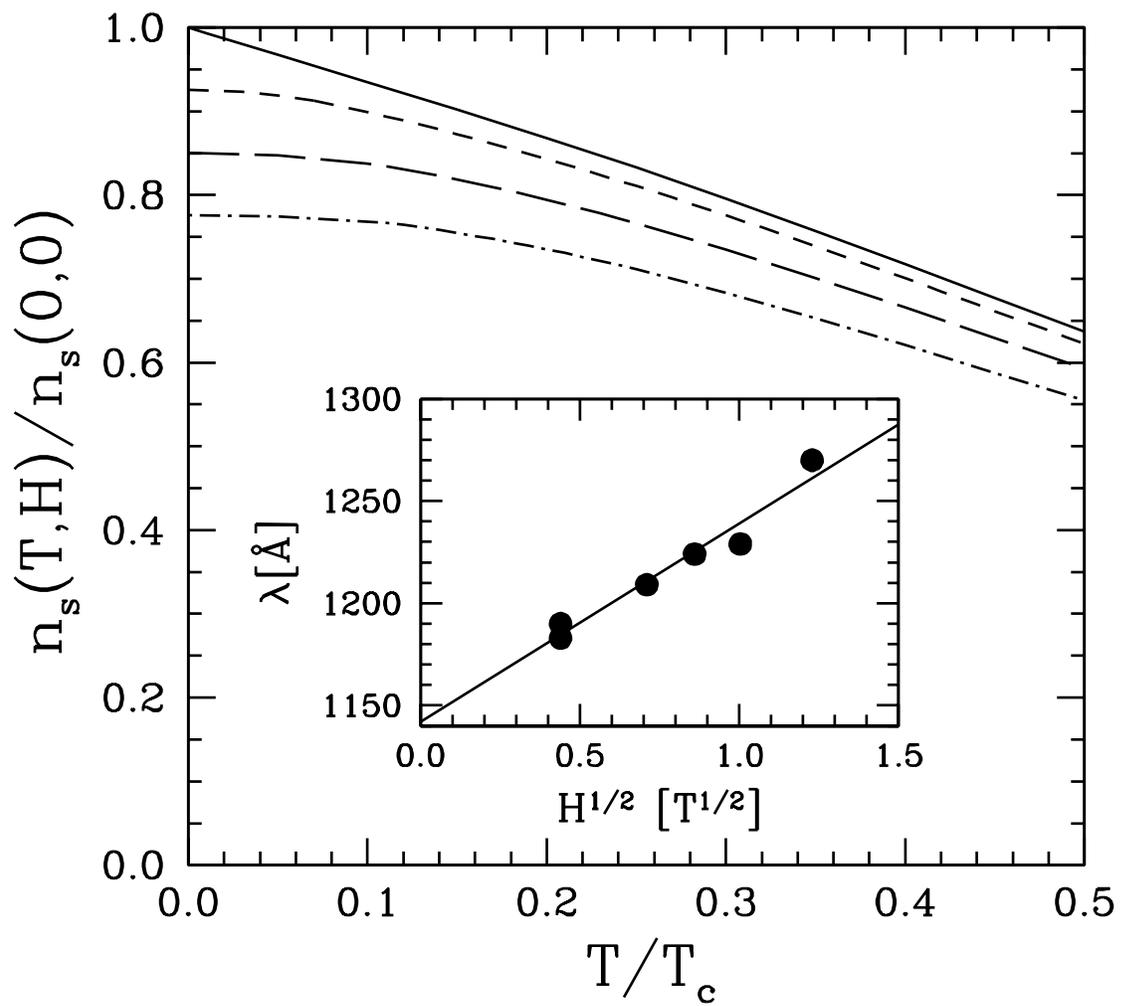
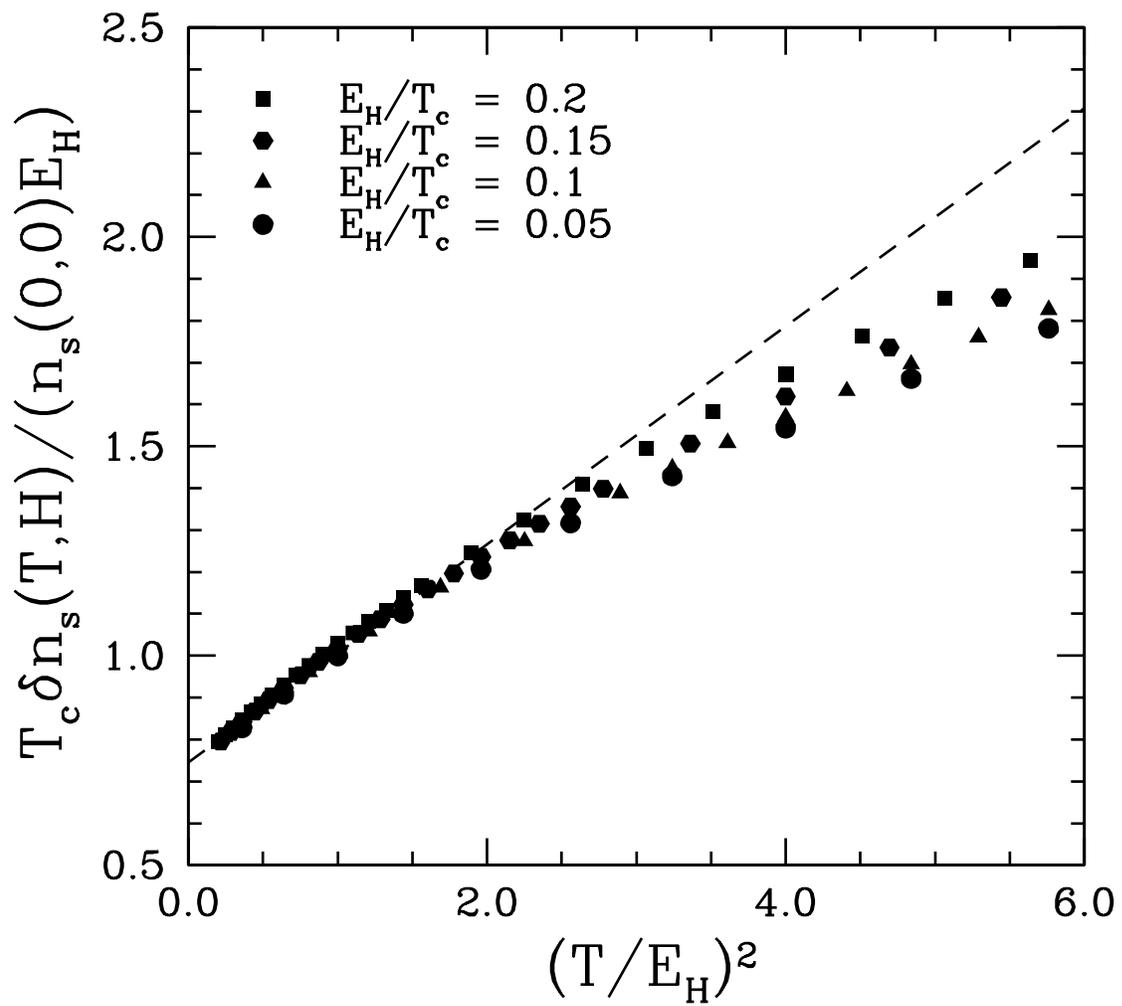
\end{document}